\documentclass{article}

\usepackage{amsmath}
\usepackage{amsfonts}
\usepackage{amssymb}
\usepackage{amsthm}

\usepackage{hyperref}

\usepackage{graphicx}
\graphicspath{{fig/}}

\usepackage{tikz}

\newtheorem{thm}{Theorem}

\newtheorem{lem}[thm]{Lemma}
\newtheorem{cor}[thm]{Corollary}
\newtheorem*{rem}{Remark}

\newcommand{\ds}{\text{d}s}
\newcommand{\dtau}{\text{d}\tau}
\newcommand{\dt}{\text{d}t}

\newcommand{\nono}{\nonumber}
\newcommand{\obs}{(\ref{obs2eqn}-\ref{obs_data})}
\newcommand{\obss}{(\ref{obs2eqn}-\ref{obs_data}) }
\newcommand{\be}{\beta_e}
\newcommand{\bb}{\overline{\be}}

\newcommand{\Rn}{\mathcal{R}_0}
\newcommand{\Rt}{\mathcal{R}_t}
\newcommand{\alp}[1]{\alpha_{#1}}
\newcommand{\sig}[1]{\sigma_{#1}}

\newcommand{\F}{\mathcal{F}}

\newcommand{\eps}{\varepsilon}

\begin{document}

\title{Reformulating the SIR model in terms of the number of COVID-19 detected cases: well-posedness of the observational model}%Identification of parameters for SIR-type models from the number of active cases
\author{Eduard Campillo-Funollet$^1$, \\Hayley Wragg$^2$, \\James Van Yperen$^2$, \\Duc-Lam Duong$^2$, \\Anotida Madzvamuse$^2$}

%\subject{Mathematics, Infectious Diseases}

%\keywords{SIR, Epidemiology, Existence, Uniqueness, Observational Model}

%\corres{Eduard Campillo-Funollet \\
%\email{e.campillo-funollet@kent.ac.uk} \\
%Anotida Madzvamuse \\
%\email{a.madzvamuse@sussex.ac.uk}}

%\begin{fmtext}
%Not entirely sure what to put here, I think it is anything extra that we need on the front-page
%\end{fmtext}

\maketitle
\begin{abstract}
    Compartmental models are popular in the mathematics of epidemiology for their simplicity and wide range of applications. Although they are typically solved as initial value problems for a system of ordinary differential equations, the observed data is typically akin of a boundary value type problem: we observe some of the dependent variables at given times, but we do not know the initial conditions. In this paper, we reformulate the classical Susceptible-Infectious-Recovered system in terms of the number of detected positive infected cases at different times, we then prove the existence and uniqueness of a solution to the derived boundary value problem and then present a numerical algorithm to approximate the solution. 
\end{abstract}

{$^1$School of Mathematics, Statistics and Actuarial Science, Department of Statistical Methodology and Applications, University of Kent, CT2 7PE, UK\\ 
$^2$School of Mathematical and Physical Sciences, Department of Mathematics, University of Sussex, BN1 9QH, UK}

\section{Introduction}

Since the onset of COVID-19, the disease caused by SARS-CoV-2, in December 2019 and the declaration of a pandemic by the World Health Organisation in March 2020, governments and organisations across the globe have implemented unprecedented regional, national and international interventions to contain the spread of COVID-19 and the subsequent damage caused by it. Initially non-pharmaceutical interventions (NPIs) were the only tool in the fight to contain the spread, forcing governments to introduce strict control measures on public mixing, the use of personal protective equipment, improvement of personal hygiene and the use of contact tracing to reduce the possibility of transmission throughout their countries, protecting those most vulnerable as well as maintaining control of the healthcare demand and capacity. The social-economical effects these necessary decisions are still unknown and it is speculated that these effects will still be felt for many more years to come \cite{DKTMM20,CGK20,PN20,NASal20}. The global picture is starting to look less bleak with the introduction of effective vaccines and vaccination policies allowing the relaxation of such strict NPIs, however with the spawn of new variants of COVID-19, such as the alpha variant (lineage B.1.1.7) and more recently the delta variant (lineage B.1.617.2), there is still plenty of work to do \cite{HFJK21,HHZADD21,DABal21,MHTDK21}. Indeed, now that "normality" is returning, country borders are opening and tourism is on the increase, we in the UK are starting to see an increase in daily positive cases, hospitalisations and deaths. 

At the beginning of the pandemic, mathematical modelling of infectious diseases was thrust into the limelight for governments across the globe to gain imperative forecasting pictures of the impending spread of COVID-19, and NPIs were interpreted by adjusting parameters to allow for quick decision-making on the appropriate NPI measures to be implemented. At the forefront of these epidemiological models that have played a pivotal role throughout the pandemic is the Imperial College Model \cite{FLNal20} which set the precedent on the forecasting capabilities of standard Susceptible-Infected-Removed (SIR) models for COVID-19. At its inception, lots of assumptions had to be made about various epidemiological characteristics, such as incubation time and latency, of COVID-19 either using datasets from China or using past information on similar infectious diseases such as Influenza or SARS. The bonus of using a mathematical model in this setting is that, upon gaining new information and data, it is fairly standard to update the model to include this new information. Since then, lots of other models answering questions about COVID-19 have been created pushing the frontier of the field, such as \cite{CPKal21, KSMal21, OSAal20, HPNal21, MHG21}. However models of grandeur like the Imperial College Model are difficult to validate due to the number of assumptions, the nature of the assumptions, and the lack of data available to cross-reference against the model outputs \cite{CVAal21,BFMSS20,FWKal20,SBBal20,JLJ20}. This opens up two questions: the question of modelling approaches and the question of data availability and parameter identification. One can argue that having a very detailed model allows for flexibility of forecasting a large range of potential scenarios, however if the results can not be validated then how can the results be relevant. On the other hand, using a simpler model may allow for validation against data but has little forecasting power. Taking a data-driven approach to the derivation of a model, using a principle like Occam's razor, is becoming popular amongst the community, but requires good quality data \cite{OSAal20,CVAal21,BFMSS20,SBBal20,JLJ20,ABDWW06,A20}. The global movement for setting up dashboards and repositories where COVID-19 data is publicly accessible has dramatically improved the ability to progress mathematical modelling, such as in \cite{UKgov,JHdash}. The data in these types of repositories is typically comprised of positive cases, vaccination status, hospitalisations and deaths, ranging from nationally to locally. The granularity of this data has improved over the course of the pandemic, allowing for the appearance of age-structured models and models that are able to differentiate between the medically vulnerable and otherwise. In particular, in this publication we focus on the positive cases and testing throughout the globe. The testing strategy and reporting has vastly increased in precision throughout the pandemic to allow for reporting on the approximate number of individuals infectious in the community on any one day - this number is often the basis of the dynamics of mathematical models of infectious diseases, especially when it comes to population dynamics models like the SIR model. Understanding and predicting levels of infections in the community is the crux of mathematical modelling in epidemiology and also, arguably, the hardest to quantify. For example, a major difficulty with COVID-19 is that a large majority of infected individuals are asymptomatic whilst those who seek testing are typically symptomatic. Similarly, the initial display of symptoms is not necessarily at the same time an individual becomes infectious. Testing for both of these cases is tricky without constant mass testing, so there is already a level of uncertainty before any parameter estimation can be done. This often leads to using a parameter which describes under-reporting. With SIR type models, this has an effect on the fitted transmission rates which are often the main parameter to be manipulated when NPIs are modelled. 

Not only is lack of data problematic when deriving and using SIR type mathematical models, the context and interpretation of the data and the parameter fitting process often comes with difficulties. In particular, there is often not a one-to-one correspondence between data and direct parameter or compartment in a model, especially when considering daily snapshots of an epidemic for example. This is the main motivation behind the so-called "observational model", a representation of an SIR type model described only in terms of the model parameters and compartments that are captured by the mathematical interpretation of the data \cite{CVAal21}. This observational model gives us an intuitive understanding of what parameters can be identified and how the data affects different compartments in the model. A standard procedure is to treat the mathematical model as an initial value problem and to either fit or seed when the initial infection might have been \cite{MHTDK21,FLNal20,KSMal21,CVAal21,KTGGL20,CWELBP21}. What we argue in this paper is that, using the data available, one can in fact treat the problem as a boundary value problem and that the formulation of the observational model as a boundary value problem is well-posed. What is still to be explored is how one can use the boundary value problem to conduct parameter estimation and to deal with noise within the data, but we leave this for a future publication. 

The structure of the paper is as follows: in Section \ref{Obs_model} we derive the observational model and pose the boundary value problem, in Section \ref{existence} we prove existence of a solution to the boundary value problem, in Section \ref{uniqueness} we prove uniqueness of the solution to the boundary value problem and in Section \ref{numerics} we outline an approach to numerically approximate the solution to the boundary value problem. 

\section{Derivation of the Observational Model}
\label{Obs_model}
\begin{figure}[!ht]
    \centering
    \begin{tikzpicture}[cap=round,>=latex]
    \tikzstyle{comp} = [circle, minimum size=1cm, draw=black, text centered]
    \tikzstyle{compr} = [circle, minimum size=1cm, draw=red, text centered, line width=0.75mm,dashed]
    \tikzstyle{arrow} = [thick,->,>=stealth]
    \tikzstyle{arrowb} = [ultra thick,dotted,->,>=stealth,blue]
    \tikzstyle{arrowg} = [ultra thick,dashed,->,>=stealth,red]
    \node (S) [comp] {$S$};
    \node (I) [comp, right of=S, xshift=1.5cm] {$I$};
    \node (R) [comp, right of=I, xshift=1.5cm] {$R$};
    \draw [arrow] (S) to node[anchor=south] {$\lambda(t)$} (I);
    \draw [arrow] (I) to node[anchor=south] {$\gamma$} (R);
    \end{tikzpicture}
    \caption{Schematic representation of the SIR compartmental model.}
    \label{fig:model}
\end{figure}
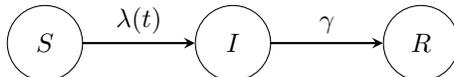

We begin by introducing the standard SIR equations for compartmental models, often accredited to the seminal work of Kermack and McKendrick in 1927 \cite{KM27}; see Figure \ref{fig:model} for a schematic representation. The mathematical model takes the following form of a simple temporal epidemiological dynamic system of ordinary differential equations, supported by non-negative initial conditions
\begin{align}
    \dot{S} &= - \beta \frac{I}{N} S, \quad & \, S(0) = S_0, \label{Seqn} \\
    \dot{I} &= \beta \frac{I}{N} S - \gamma I, \quad & I(0) = I_0, \label{Ieqn} \\
    \dot{R} &= \gamma I, \quad & R(0) = R_0. \label{Reqn}
\end{align}
Here, the dot above the notation denotes the time derivative. In this setting, let $N$ denote the total population being considered. Then, $S(t)$ denotes the proportion of the total population $N$ who are susceptible to the infectious disease being studied. Susceptible individuals become infectious with the disease to form the $I(t)$ subpopulation at a rate $\lambda(t)$ which represents the current infection rate. The rate $\lambda(t)$ is the product between $\beta$, the average transmission rate, and the probability of meeting an infectious person $I(t)N^{-1}$. Individuals in the $I(t)$ subpopulation then are removed at a rate $\gamma$ to form the removed subpopulation $R(t)$. The description of the removed compartment depends on the nature of the infectious disease and the interpretation of the data available to be used. Typically the term removed is interchanged with the term recovered. As is standard for epidemiological models of this nature, $\beta^{-1}$ denotes the average time between transmissions and $\gamma^{-1}$ denotes the average length of time before removal. The initial conditions are often chosen so that $S_0 + I_0 + R_0 = N$, i.e. everyone in the population is accounted for and is in one of the compartments. This gives rise to one of many properties of the SIR equations, namely the conservation of population $S(t) + I(t) + R(t) = N$ for any $t \in [0,\infty)$, which can be shown by adding equations \eqref{Seqn}--\eqref{Reqn} and integrating in time. 

Given that the SIR equations describe the dynamics of infectious diseases, one may ask the question: how can we know from the parameters that there is going to be an epidemic? This question is most often answered by looking at when the infectious subpopulation is increasing, which leads mathematicians to define the basic reproduction number, often denoted by $\Rn$, and the effective reproduction number, often denoted by $\Rt$ and referred to as the ``R" number, which, in the case of \eqref{Seqn}--\eqref{Reqn}, take the form
\begin{align*}
    \Rn := \frac{\beta}{\gamma} \quad \text{ and } \quad \Rt := \mathcal{R}_0 \frac{S(t)}{N} %\label{R0}
\end{align*}
respectively. The parameter $\Rn$ is a bifurcation parameter such that if $\Rn < 1$ then there is no increase in the number of infectious individuals, and if $\Rn > 1$ we experience an epidemic. The parameter $\Rn$ also enables modellers to understand the initial rate of exponential increase in the number of infectious individuals since, by considering small $t$, we can informally set $S(t) \approx N$ and find that
\begin{align*}
    \dot{I} \approx (\beta - \gamma) I \quad \Longleftrightarrow \quad I(t) \approx I_0 \, e^{\gamma(\Rn - 1)t}. %\label{exponential_inc}
\end{align*}

Before deriving the observational model, we need to discuss and interpret the data in terms of the model parameters and compartments. Typically data for infectious diseases comes as a daily snapshot of the pathway of how individuals experience the infectious disease, such as deaths or hospitalisations. In this study we are looking to use the detected cases of an infectious disease. In order to describe this in terms of the model parameters and compartments, we formally describe the removed compartment $R$ as those who are self-isolating and thus are not mixing in the general population. Hence, for each observation of the number of detected cases $X_m$, $m = 0, \dots, M$, we see that
\begin{align}
    X_m := r \gamma \int_{t_m}^{t_{m+1}} I(s) \, \ds, \label{data}
\end{align}
where $[t_m,t_{m+1}]$ represents the time-interval of the measurement of the data and $r$ denotes the proportion of under-reporting. The parameter $r$ represents a few different instances of individual behaviour, for example it could be those who do not self-isolate after a positive test, or those who do not report the result after taking a test. We note that since the data describes the change of positive cases over a certain time period, it is not clear what the initial conditions for \eqref{Seqn}--\eqref{Reqn} should be in order for us to utilise this data. With interpretation in mind, we formally state that, for any $m = 1,\dots,M$,
\begin{align*}
    %\label{data_stability}
    0 < X_m < N,
\end{align*}
that is to say that we always observe at least one positive case per time interval and that it is not possible to observe more than the whole population to test positive at the same time. 

As described in \cite{CVAal21}, since the data is described by the infectious compartment $I$, we need to re-write the equations \eqref{Seqn} and \eqref{Ieqn} purely in terms of $I$. We note that due to the conservation of population property, we in fact do not need to consider \eqref{Reqn} as $R(t) = N - (S(t) + I(t))$. In order to derive the observational model, we note three properties of equations \eqref{Seqn} and \eqref{Ieqn}, namely we have
\begin{align}
    \dot{S} + \dot{I} = -\gamma I, \label{S'I'eqn}
\end{align}
which, by differentiating, immediately gives us
\begin{align}
    \ddot{S} + \ddot{I} = -\gamma \dot{I}. \label{S''I''eqn}
\end{align}
Moreover, by differentiating \eqref{Seqn}, we see that
\begin{align}
    \ddot{S} = \dot{S} \left( \frac{\dot{I}}{I} - \beta \frac{I}{N} \right) \label{S''eqn}
\end{align}
and hence, inserting \eqref{S'I'eqn} and \eqref{S''I''eqn} into \eqref{S''eqn}, we see that the observational model takes the form
\begin{align}
    \ddot{I} = \dot{I} \left( \frac{\dot{I}}{I} - \beta \frac{I}{N} \right) - \beta \, \gamma \, \frac{I^2}{N}. \label{obs1eqn}
\end{align}
One notes that \eqref{obs1eqn} is well-defined for $t \geq 0$ since, by considering \eqref{Ieqn}, $\dot{I} I^{-1}$ is bounded. For the purposes of this paper, we continue to transform \eqref{obs1eqn} to facilitate the analysis. By denoting $\be := r^{-1} \, \beta \, N^{-1}$ and setting $z(t) = \ln(r \, \gamma \, I(t))$, \eqref{obs1eqn} is equivalent to
\begin{align}
    \ddot{z} = -\be e^z \left( \frac{\dot{z}}{\gamma} + 1 \right), \label{obs2eqn}
\end{align}
whilst the data \eqref{data} is equivalent to 
\begin{align}
    X_m = \int_{t_m}^{t_{m+1}} e^{z(s)} \, \ds. \label{obs_data}
\end{align}
The observational model \obss describes a second-order nonlinear boundary value problem with novel non-local boundary conditions. 

\begin{rem}
One can go further to simplify \eqref{obs2eqn} by taking $w(\xi) = z(t) + \ln(\be \gamma^{-2})$, where $\xi = \gamma t$, to get
\begin{align*}
    w'' = -e^w \left(w' + 1 \right) %\label{obsweqn}
\end{align*}
where the prime denotes the derivative with respect to $\xi$, which leads to \eqref{obs_data} satisfying
\begin{align*}
    X_m = \frac{\gamma}{\be} \int_{\gamma t_m}^{\gamma t_{m+1}} e^{w(\eta)} \, \text{d}\eta. %label{obs_wdata}
\end{align*}
\end{rem} 

In this publication we look to prove the following two theorems.
\begin{thm}[Existence] \label{thm:existence}
    Given $X_0,X_1>0$, corresponding to \eqref{obs_data}, and given parameters $\beta_e,\gamma > 0$, there exists a solution $z\in C^2(t_0,t_2)$ to \eqref{obs2eqn} subject to \eqref{obs_data}. 
\end{thm}
\begin{thm}[Uniqueness] \label{thm:uniqueness}
    Given $X_0, X_1 > 0$ corresponding to \eqref{obs_data}, and given parameters $\beta_e,\gamma > 0$, the solution to \obss is unique in $C^2(t_0,t_2)$. Furthermore, the solution is smooth and globally unique.
\end{thm}

\section{Existence}
\label{existence}

The proof of Theorem \ref{thm:existence} is based on the Leray-Schauder fixed point theorem \cite{H08}. We shall show that the solutions to \obss satisfy \textit{a priori} bounds, and we will use them to show compactness of a suitable operator.

\begin{lem} \label{zbound_0}
Let $X_0>0$ and $X_1>0$. Suppose there exists a solution $z$ to \obss with data $X_0$ and $X_1$. Then, there exists some constant $C$ such that $|z(t)| \leq C$, for $t \in [t_0,t_2]$. Moreover, $C$ is only in terms of the parameters and the data.
\end{lem}

\proof Let $\bb := \be \, \gamma^{-1}$ and let $\tau \in [t_0,t_2]$. Integrating \eqref{obs2eqn} over $(t_{0},\tau)$ we have
\begin{align*}
    \dot{z}(\tau) - \dot{z}(t_{0}) = \bb \left( e^{z(t_{0})} - e^{z(\tau)} \right) - \be \int_{t_{0}}^\tau e^{z(s)} \, \ds. %\label{int_obs2eqn0}
\end{align*}
Let $t \in [t_{0},t_{2}]$ and let $\delta_{0}(t) := t - t_{0}$. Then, by integrating with respect to $\tau \in (t_{0},t)$, we have
\begin{align*}
    z(t) - z(t_{0}) - \delta_{0}(t) \, \dot{z}(t_{0}) = \bb \left( \delta_{0}(t) \, e^{z(t_{0})} - \int_{t_{0}}^t e^{z(\tau)} \, \dtau \right) - \be \int_{t_{0}}^t \int_{t_{0}}^\tau e^{z(s)} \, \ds \, \dtau. %\label{int2_obs2eqn0}
\end{align*}
Since the exponential function is positive everywhere, we see that
\begin{align*}
    0 \leq \int_{t_{0}}^t e^{z(\tau)} \, \dtau \leq \int_{t_{0}}^{t_{2}} e^{z(\tau)} \, \dtau = X_0+X_1, %\label{data_bound1}
\end{align*}
and consequently
\begin{align*}
    0 \leq \int_{t_{0}}^t \int_{t_{0}}^\tau e^{z(s)} \, \ds \, \dtau \leq \delta_{0}(t) \, (X_0+X_1). %\label{data_bound2}
\end{align*}
Hence, we see that there exist bounds 
\begin{align} \label{zlzu}
    z_{0,l}(t) \leq z(t) \leq z_{0,u}(t),
\end{align}
for $t\in [t_0,t_2]$, where
\begin{align}
    z_{0,l}(t) := z(t_{0}) + \delta_{0}(t) \left( \dot{z}(t_{0}) + \bb e^{z(t_{0})} \right) - \bb(1 + \gamma \, \delta_{0}(t)) (X_0+X_1), \label{zl0}
\end{align}
and 
\begin{align}
    z_{0,u}(t) := z(t_{0}) + \delta_{0}(t) \left( \dot{z}(t_{0}) + \bb e^{z(t_{0})} \right). \label{zu0}
\end{align}
These bounds depend on $z$ and $\dot{z}$ through their values at $t_0$. Since we are interested in uniform bounds for any solution $z$, we need to show that $z(t_0)$ and $\dot{z}(t_0)$ are bounded in terms of the data $X_0$ and $X_1$. By exponentiation and integration \eqref{zlzu} over $(t_0,t_2)$, we have
\begin{align}
    \int_{t_{0}}^{t_{2}} e^{z_{0,l}(t)} \, \dt \leq (X_0+X_1) \leq \int_{t_{0}}^{t_{2}} e^{z_{0,u}(t)} \, \dt, \label{int3_obs2eqn0}
\end{align}
and hence, by taking \eqref{zl0} into account,
\begin{align*}
    \int_{t_{0}}^{t_{2}} e^{z_{0,l}(t)} \, \dt &= e^{z(t_{0}) - \bb (X_0+X_1)} \int_{t_{0}}^{t_{2}} e^{\delta_{0}(t) \left( \dot{z}(t_{0}) + \bb e^{z(t_{0})} - \be (X_0+X_1) \right)} \, \dt  \nono \\
        &= \frac{e^{z(t_{0}) - \bb (X_0+X_1)}}{\dot{z}(t_{0}) + \bb e^{z(t_{0})} - \be (X_0+X_1)} \left( e^{\delta_0(t_2) \left( \dot{z}(t_{0}) + \bb e^{z(t_{0})} - \be (X_0+X_1) \right)} - 1 \right). %\label{ezl0}
\end{align*}
To show that $z(t_0)$ and $\dot{z}(t_0)$ are bounded, we proceed by contradiction. Let $\alp{0} := e^{z(t_{0})}$ and $\sig{0} := \dot{z}(t_{0})$, and let
\begin{align*}
    f_{0,l}(\alp{0},\sig{0}) := \frac{\alp{0} \, e^{-\bb(X_0+X_1)}}{\sig{0} + \bb \alp{0} - \be(X_0+X_1)} \left(e^{\delta_0(t_2) \left(\sig{0} + \bb \alp{0} - \be(X_0 + X_1) \right)} - 1 \right). %\label{flzl0}
\end{align*}
A standard application of the L'H\^{o}pital's rule yields
\begin{align*}
    \lim_{\alp{0} \rightarrow \infty} f_{0,l}(\alp{0},\sig{0}) = \infty \quad \text{ and } \quad \lim_{\sig{0} \rightarrow \infty} f_{0,l}(\alp{0},\sig{0}) = \infty, %\label{limflzl0}
\end{align*}
where we have fixed $\sig{0}$ in the first limit and fixed $\alp{0}$ in the second limit, and similarly
\begin{align*}
    \lim_{\alpha_0,\sigma_0 \rightarrow \infty} f_{0,l} = \infty. %\label{limflzl0_dbl}
\end{align*}
Since \eqref{int3_obs2eqn0} implies boundedness of $f_{0,l}$, we conclude $\alpha_0$ and $\sigma_0$ cannot be arbitrarily large and in consequence, $z(t_0)$ and $\dot{z}(t_0)$ are bounded from above. Using a similar argument with the right hand side bound of \eqref{int3_obs2eqn0} and taking the limits $\alpha_0 \rightarrow 0$ and $\sigma_0 \rightarrow -\infty$, we can conclude that $z(t_0)$ and $\dot{z}(t_0)$ are also bounded from below.

Therefore, the bounds \eqref{zl0} and \eqref{zu0} are independent of the solution $z$, and depend only and the parameters and the data.\qed \\

We can now proceed to show that $\dot{z}$ is also uniformly bounded in $[t_0,t_2]$.

\begin{lem} \label{z'bound}
Let $X_0>0$ and $X_1>0$. Suppose there exists a solution $z$ to \obss with data $X_0$ and $X_1$. Then, there exists some constant $C'$ such that $|\dot{z}(t)| \leq C'$, for $t \in [t_{0},t_{2}]$. Moreover, $C'$ is only in terms of the parameters and the data. 
\end{lem}

\proof Using the notation introduced in the proof of Lemma \ref{zbound_0}, we integrate \eqref{obs2eqn} over $(t_0,t)$, where $t \in [t_0,t_2]$, to obtain 
\begin{align}
    \dot{z}(t) = \dot{z}(t_0) -  \bb \left( e^{z(t)} - e^{z(t_0)} \right) - \be \int_{t_0}^t e^{z(s)} \, \ds. \label{eqn:int_obs2eqn}
\end{align}
We can now use the uniform bounds on $z$, from Lemma \ref{zbound_0}, together with the bounds on $z(t_0)$ and $\dot{z}(t_0)$ that we obtained in the proof of Lemma \ref{zbound_0}, to bound the right hand side of \eqref{eqn:int_obs2eqn}.\qed \\

We can proceed using equation \obss and the bounds on $z$ and $\dot{z}$ to obtain uniform bounds on the second derivative, which in turns allows us to obtain uniform bounds for the third derivative by using them in \eqref{obs2eqn}. Indeed, we can bootstrap this argument to obtain uniform bounds up to the $n$-th derivative, for finite $n$. We summarise this in the following corollary.

\begin{cor} \label{thm:all_bounds}
Let $n=0,1,\dots,\overline{N}$, with $\overline{N} < \infty$, and let $X_0>0$ and $X_1>0$. Suppose there exists a solution $z$ to \obss with data $X_0$ and $X_1$. Then, there exists some constant $C^{(n)}$, depending only on $n$, the data and the parameters, such that $|z^{(n)}(t)| < C^{(n)}$ for all $t\in[t_0,t_2]$.
\end{cor}

We can now use the uniform bounds on the solutions to study a suitable operator that will allow us to define the solutions of \obss as a fixed point. Let
\begin{align}
    \F[z](t) := a_z + b_z t - \int_{t_0}^{t} 
\int_{t_0}^\tau \be \, e^{z(s)} \left( \frac{\dot{z}(s)}{\gamma} + 1 \right) \, \ds \, \dtau, \label{F_operator}
\end{align}
where $a_z,b_z$ depend on $z$ and are uniquely determined under the constraints
\begin{align}
    \int_{t_0}^{t_1} e^{\F[z](t)} \, \dt = X_0,\quad \int_{t_1}^{t_2} e^{\F[z](t)} \, \dt = X_1. \label{F_operator_cons}
\end{align}
We note importantly that a fixed point of $\F$ is a solution to \obs. Therefore, our goal is to apply the Leray-Schauder fixed point theorem to $\F$ to show the existence of a solution in $C^2(t_0,t_2)$. We will first show that $\F$ is a compact operator. 

\begin{lem} \label{compactness}
    Let $\F$ be defined as in \eqref{F_operator}. Then $\F$ is a compact operator in $C^2(t_0,t_2)$. 
\end{lem}

\proof Let $(z_n)_{n \in \mathbb{N}_0}$ be a sequence of functions bounded in $C^2(t_0,t_2)$. We see that since
\begin{align*}
    (\F[z_n])^{(3)}=- \be \, \dot{z}_n \, e^{z_n}\left(\frac{\dot{z}_n}{\gamma} + 1\right) - \be e^{z_n} \frac{\ddot{z}_n}{\gamma}, %\label{F'''}
\end{align*}
we have $(\F[z_n])^{(3)}$ is also uniformly bounded, since $(z_n)_n$ is uniformly bounded in $C^2(t_0,t_2)$. Therefore, the Arzela-Ascoli theorem implies that $\F$ is a compact operator in $C^2(t_0,t_2)$. We note that to apply the Arzela-Ascoli theorem in $C^2(t_0,t_2)$ suffices to show uniform boundedness of the third derivative, since then we also have equicontinuity of the second derivative; this is analogous to using uniform boundedness of the first derivative to show equicontinuity of the sequence of functions. \qed \\

With this set up, in particular that $\F$ is compact in $C^2(t_0,t_2)$, we can now finish the proof of Theorem \ref{thm:existence}. \\

\noindent \textit{Proof of Theorem \ref{thm:existence}}. In order to apply the Leray-Schauder fixed point theory, we now need to show that the set
\begin{align}
    \left\{ z \in C^2(t_0,t_2) \text{ : } z = \kappa \F[z] \right\} \label{eqn:ls_set}
\end{align}
is bounded for all $\kappa \in[0,1]$. It is easy to see, by linearity of the derivative and the integral, that a fixed point of $\kappa \F$ corresponds to a fixed point of $\F$ with a change of parameters, namely, by replacing $\be$ by $\kappa \be$. Therefore, any function $z$ in \eqref{eqn:ls_set} is a solution to \obss with a suitable parameter, and as shown in Lemma \ref{zbound_0}, is bounded in $C^2(t_0,t_2)$. Thus Leray-Schauder fixed point theorem implies the existence of a fixed point for $\F$, which, as already mentioned, corresponds to a solution of \obs. \qed

\section{Uniqueness} 
\label{uniqueness}

Now that we have proven that there exists a solution to \obs, we now want to demonstrate the uniqueness of a solution to \obs. We employ a standard contradiction argument for boundary value problems by showing that two different solutions, which satisfy the same boundary conditions, must be the same. \\

\noindent \textit{Proof of Theorem \ref{thm:uniqueness}.} Let $x(t)$ and $y(t)$ be two solutions to \obss satisfying the same data conditions, that is for $m = 0, 1$ we have
\begin{align*}
    X_m = \int_{t_m}^{t_{m+1}} e^{x(s)} \, \ds = \int_{t_m}^{t_{m+1}} e^{y(s)} \, \ds. %\label{uniq_obs_data}
\end{align*}
Consequently, this means that 
\begin{align}
    \int_{t_m}^{t_{m+1}} e^{x(s)} - e^{y(s)} \, \ds = 0. \label{E0}
\end{align}
Let $D(t) := x(t) - y(t)$ and $E(t) := e^{x(t)} - e^{y(t)}$. By taking the difference of \eqref{obs2eqn} for each solution, we have
\begin{align}
    \ddot{D} = -\be \left(E + \frac{\dot{E}}{\gamma} \right). \label{eqn:diff_solutions}
\end{align}
To simplify the presentation, we set $\bb := \be \, \gamma^{-1}$ and denote 
\begin{align*}
    F(t) = \dot{D}(t) + \bb E(t) - \dot{D}(t_0) - \bb E(t_0). %\label{F_notation}
\end{align*}
Using this notation, \eqref{eqn:diff_solutions} is equivalent to
\begin{align}
    \label{eqn:F'diff_solutions}
    \dot{F} + \be E = 0.
\end{align}
Let $t \in [t_0,t_2]$. Integrating \eqref{eqn:F'diff_solutions} with respect to $s \in (t_0,t)$ gives
\begin{align}
    \label{eqn:F_diff_solutions}
    F(t) + \be \int_{t_0}^t E(s) \, \ds = 0.
\end{align}
Moreover, by \eqref{E0} and \eqref{eqn:F_diff_solutions}, one can see that 
\begin{align}
    F(t_1) = F(t_2) = 0. \label{F0}
\end{align}
We now proceed by contradiction to show that $D(t_0) = 0$. Without loss of generality, assume that $D(t_0) > 0$. This gives us that $E(t_0) > 0$ by definition and $\dot{F}(t_0) < 0$ by \eqref{eqn:F'diff_solutions}. Considering this, \eqref{F0} and the fact that $x$, $y \in C^2(t_0,t_2)$, we have that there exists $\tau \in (t_0,t_1]$ such that $F(\tau) = 0$ and $F(t) < 0 $ for all $t \in (t_0,\tau)$. In particular, this means that, by integrating $F$ with respect to $s \in (t_0,\tau)$, we have
\begin{align}
    \label{eqn:tbar_F}
    \int_{t_0}^{\tau} F(s) \, \ds < 0
\end{align}
and, setting $t = \tau$ in \eqref{eqn:F_diff_solutions}, we have
\begin{align}
    \label{eqn:tbar_E}
    \int_{t_0}^{\tau} E(s) \, \ds = 0.
\end{align}
Then, by integrating \eqref{eqn:F_diff_solutions} with respect to $t \in (t_0,\tau)$ and using \eqref{eqn:tbar_F}, we obtain
\begin{align*}
    0 = \int_{t_0}^{\tau} F(t) \, \dt + \be \int_{t_0}^{\tau}\int_{t_0}^{t} E(s) \, \ds \, \dt < \be \int_{t_0}^{\tau}\int_{t_0}^{t} E(s) \, \ds \, \dt = 0, %\label{eqn:contradiction}
\end{align*}
where the last equality is a consequence of Fubini's theorem and \eqref{eqn:tbar_E}. Therefore we have a contradiction. One notices that, using the same approach, we can also conclude that $D(t_1) = 0$ due to \eqref{F0}. By integrating \eqref{eqn:F_diff_solutions} with respect to $t \in (t_0,t_1)$, we have that
\begin{align*}
    D(t_1) + \bb \int_{t_0}^{t_1} E(t) \, \dt + \be \int_{t_0}^{t_1} \int_{t_0}^{t} E(s) \, \ds \, \dt = D(t_0) + (t_1 - t_0) \left( \dot{D}(t_0) + \bb E(t_0) \right), %label{int_F}
\end{align*}
which, by noting that $D(t_0) = 0$ implies that $E(t_0) = 0$ and using \eqref{E0}, gives us
\begin{align*}
    \dot{D}(t_0) = 0. %\label{D'0}
\end{align*}
Standard ODE theory implies that, since $x(t_0) = y(t_0)$, $\dot{x}(t_0) = \dot{y}(t_0)$ and $x$, $y \in C^2(t_0,t_2)$, we have $x(t) = y(t)$ for all $t \in [t_0,t_2]$.  \qed 

\section{Numerical Example}
\label{numerics}

\begin{figure}[!ht]
    \centering
    \includegraphics[width=3.25in]{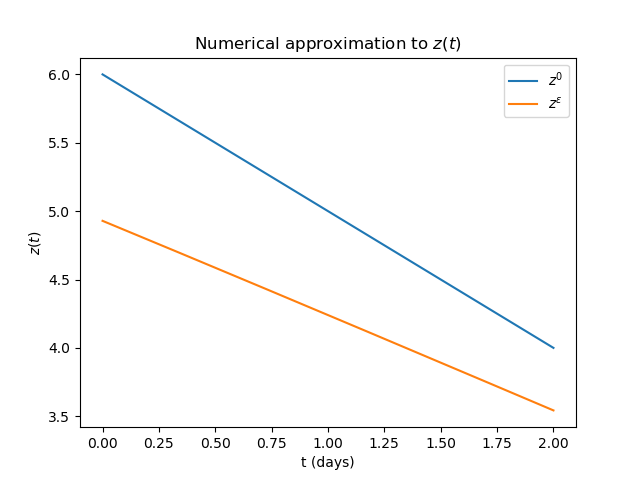}
    \caption{Comparison of the initial guess for $z$ and the converged numerical approximation for $z^\eps$.}
    \label{fig_sim}
\end{figure}

Now that we have demonstrated existence and uniqueness, we can build a numerical algorithm to approximate the solution to \obs. We set up the numerical problem as follows: given $X_0$, $X_1$, $\be$ and $\gamma$, we look to find the function $z$ which satisfies \obs. Standard numerical methods will not work in this setting due to the combination of novel non-local boundary conditions \eqref{obs_data} which depend on the solution to the second-order nonlinear differential equation \eqref{obs2eqn}. Typically, problems of this caliber are solved using an iterative approach. We start by making a guess at the initial conditions of $z$ and $\dot{z}$, which we denote by a superscript 0. The algorithm we employ returns the initial conditions which satisfy \eqref{obs_data} up to some tolerance and we use those initial conditions to reconstruct the converged solution to \obs, which we denote by a superscript $\eps$. We use Newton's method to provide the iterative approach to get the correct initial conditions and, for each iteration of the Newton's algorithm, we use a standard numerical method for the initial value problem to approximate the solution to \eqref{obs2eqn}, see for example \cite{PVTF86} for details of the individual numerical methods. As a demonstration, we set $\be = 3.75\times10^{-4}$, $\gamma = 1$ and use the data $X_0 = 100$ and $X_1 = 50$ with $t_0 = 0$, $t_1 = 1$ and $t_2 = 2$. We use an initial guess of $z^0(0) = 6$, $\dot{z}^0(0) = -1$, depicted by the blue line in Figure \ref{fig_sim}, which results in
\begin{align*}
    \int_0^1 e^{z^0(s)} \, \ds \approx 255.00079468571883 \quad \text{ and } \quad \int_1^2 e^{z^0(s)} \, \ds \approx 93.80954898172891.
\end{align*}
Using the algorithm described above, the resulting initial conditions are $z^\eps(0) = 4.92905696$ and $\dot{z}^\eps(0) = -0.68242124$, with the resulting function $z^\eps$ depicted in orange in Figure \ref{fig_sim}. 

\begin{figure}[!ht]
    \centering
    \includegraphics[width=3.25in]{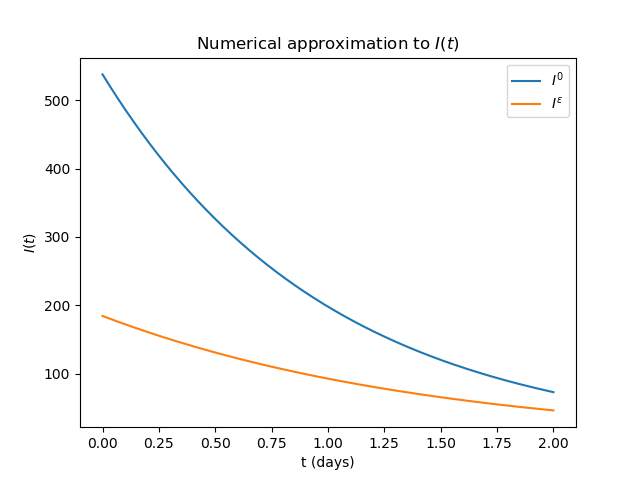}
    \caption{Comparison of the transformed initial guess for $I$ and the transformed converged numerical approximation for $I^\eps$.}
    \label{fig_sim2}
\end{figure}

Even though we can gain knowledge of the total population $N$, we can not extract $\beta$ from $\be$ due to the under-reporting parameter $r$. However, for the purpose of demonstration, we fix $r=0.75$ to allow us to demonstrate $I^0$ and $I^\eps$, the number of infectious people transformed from $z^0$ and $z^\eps$ respectively. Figure \ref{fig_sim2} depicts the transformed initial guess $I^0$ in blue and depicts the transformed converged solution $I^\eps$ in orange. 

\section{Conclusion}
\label{conclusion} 
In this publication we have demonstrated how to derive the observational model given data that describes the infectious compartment $I$, shown that a solution to this formulation exists and that the solution is unique. We have then gone on to describe an algorithm to approximate the solution to the nonlinear observational model. 

Now that we have guaranteed that the solution to the observational model is unique, given some parameters, we next look to ask the question of the identifiability of the parameters from the data. That is to say, how much data do we need to be able to uniquely identify the parameters $\be$ and $\gamma$ as well as $z$. This is a question of particular importance as understanding what we can identify helps us to understand the scenario-based forecasting capabilities of the mathematical model. Understanding this and therefore the underlying assumptions needed to produce forecasting results, given the limelight of compartmental models in recent history, will allow for a confident and trustworthy exchange of knowledge between mathematical modellers and those with research questions, such as local government or healthcare providers. 

As a final note, the main motivation for studying this problem is due to the COVID-19 pandemic and the resulting uphill struggle for mathematical modellers. However, the concept of deriving an observational model from a model of a real-life phenomenon given observable data is something that transcends the mathematical modelling of infectious diseases.

\vskip6pt

\enlargethispage{20pt}

%\ethics{The authors declare that they have no ethical statements.}

\subsection*{Author contributions}
{ECF, JVY and AM conceived the research question. ECF, HW, JVY, DLD carried out the mathematical analysis. ECF and JVY assembled the figures and drafted the manuscript. All authors read and approved the final manuscript.}

\subsection*{Funding}
{HW was supported by the Higher Education Innovation Fund through the University of Sussex, JVY and DLD were supported by Brighton and Hove City Council, East and West Sussex County Council and the NHS Sussex Commissioners, and ECF was supported by the Wellcome Trust grant number 204833/Z/16/Z. This work was partly supported by the Global Challenges Research Fund through the Engineering and Physical Sciences Research Council grant number EP/T00410X/1: UK-Africa Postgraduate Advanced Study Institute in Mathematical Sciences (AM, ECF). AM's work was partially funded by grants from the Health Foundation (1902431) and the NIHR (NIHR133761) and by an individual grant  from the Dr Perry James (Jim) Browne Research Centre on Mathematics and its Applications (University of Sussex). AM is a Royal Society Wolfson Research Merit Award Holder funded generously by the Wolfson Foundation.}

\subsection*{Acknowledgements}
{All authors acknowledge the continued support and collaboration of Brighton and Hove City Council, East and West Sussex County Council and the NHS Sussex Commissioners and thank them for the opportunity for past collaborations which lead to the conceiving of this manuscript.}

%\disclaimer{The authors declare that they have no disclaimer text.}

\end{document}